\documentclass{article}
\usepackage[numbers]{natbib}
\usepackage[final]{neurips_2024}
\usepackage{svg}
\usepackage[utf8]{inputenc} 
\usepackage[T1]{fontenc}   
\usepackage{hyperref}    
\usepackage{url}        
\usepackage{booktabs} 
\usepackage{amsfonts} 
\usepackage{nicefrac}    
\usepackage{microtype}     
\usepackage{xcolor}        
\usepackage{graphicx}       
\usepackage{pifont}
\usepackage{booktabs}
\usepackage{multirow}
\usepackage{caption, subcaption}
\usepackage{amsmath}
\usepackage{wrapfig}
\usepackage{enumitem}
\newcommand{\cmark}{\ding{51}}%
\newcommand{\xmark}{\ding{55}}%
\newcommand{\mrhubert}{\texttt{MR-HuBERT}}
\newcommand{\hubert}{\texttt{HuBERT}}
\newcommand{\bfoura}{\texttt{B4-a}}
\newcommand{\bfivea}{\texttt{B5-a}}
\newcommand{\btwoa}{\texttt{B2-a}}
\newcommand{\btwob}{\texttt{B2-b}}
\title{An Empirical Analysis of Speech Self-Supervised Learning at Multiple Resolutions}
\author{%
  Theo Clark,  Benedetta Cevoli,  Eloy de Jong,\\
  \textbf{Timofey Abramski,  Jamie Dougherty} \\
  Speechmatics\\
  \texttt{\{theoc,benedettac\}@speechmatics.com} \\
}

\begin{document}

\maketitle

\begin{abstract}
Self-supervised learning (SSL) models have become crucial in speech processing, with recent advancements concentrating on developing architectures that capture representations across multiple timescales. The primary goal of these multi-scale architectures is to exploit the hierarchical nature of speech, where lower-resolution components aim to capture representations that align with increasingly abstract concepts (e.g., from phones to words to sentences). Although multi-scale approaches have demonstrated some improvements over single-scale models, the precise reasons for these enhancements have poor empirical support. In this study, we present an initial analysis of layer-wise representations in multi-scale architectures, with a focus on Canonical Correlation Analysis (CCA) and Mutual Information (MI). We apply this analysis to Multi-Resolution HuBERT ($\mrhubert$) and find that (1) the improved performance on SUPERB tasks is primarily due to the auxiliary low-resolution loss rather than the downsampling itself, and (2) downsampling to lower resolutions neither improves downstream performance nor correlates with higher-level information (e.g., words), though it does improve computational efficiency. These findings challenge assumptions about the multi-scale nature of $\mrhubert$ and motivate the importance of disentangling computational efficiency from learning better representations.
\end{abstract}

\section{Introduction}
Self-supervised learning (SSL) has become a cornerstone in state-of-the-art speech processing models \cite{baevski2020wav2vec,hsu2021hubert, chen2022wavlm}. These models serve as feature extractors or pre-trained encoders for various tasks, including Automatic Speech Recognition (ASR), Speaker Diarisation, Speech Enhancement, and as inputs to Large Language Models. The versatility of a single pre-trained model across multiple downstream tasks has led to concentrated efforts on improving this foundational component.

At the same time, there is growing interest in developing SSL models that more closely emulate human learning processes, as doing so could unlock more efficient and flexible learning mechanisms \cite{lecun2022path, hole2021thousand}. While significant differences exist between the human brain and deep learning models, SSL aligns with some aspects of human cognition \cite{ssl_child}. One key feature of human learning is the multi-timescale evolution of our world model \cite{caucheteux2023evidence, mounir2023streamer}, resulting in a hierarchical learning structure that is more efficient than models operating on a single timescale.

Speech presents a particularly compelling domain for investigating these ideas as it is a mature field with well-established datasets \cite{panayotov2015librispeech, Kahn2020librilight, wang2021voxpopuli, ardila2020commonvoice} and benchmarks \cite{yang2021superb} consisting of different downstream tasks that operate most naturally on varying timescales: longer audio sequences are required for tasks like language identification and speaker diarisation, in contrast to phoneme recognition. Speech also exhibits a strong and implicit natural hierarchy \cite{jackendoff2002foundationsoflanguage}: sentences comprise words, which in turn consist of phones and prosodic features.

Designing architectures that optimally exploit this inherent hierarchy could enhance representation learning efficiency. Multi-scale architectures have been proposed across various domains \cite{mounir2023streamer, chung2017hierarchical, zhou2018unet++}, including speech processing \cite{chen2023speechformer++, borsos2023audiolm, park2022multiscalespeakerdiarization, yang2023uniaudio}. These approaches typically employ modular designs, with successive modules operating at progressively lower resolutions. Recent works \cite{ chen2023speechformer++, borsos2023audiolm, yang2023uniaudio, shi2024multiresolution, cuervo2022variableratehierarchicalcpc, bhati2021segmentalcontrastivepredictivecoding} that propose multi-scale architectures for speech processing tasks indicate that increasingly low-resolution representations align with increasingly abstract speech and language components, but these claims currently have limited empirical support.

Multi-Resolution HuBERT ($\mrhubert$) \cite{shi2024multiresolution} is a multi-scale architecture that augments $\hubert$ \cite{hsu2021hubert} with a low-resolution block and an associated auxiliary loss. $\mrhubert$ shows promise across various benchmarks and its success is attributed to a multi-scale structure. By using standard representation analysis techniques to examine these claims, we evaluate whether lower-resolution representations are more correlated with higher level speech and language units in multi-scale models. Our key contributions are:

\begin{itemize}[noitemsep]
\item Lower-resolution components in $\mrhubert$ models do not, as initially hypothesised, capture representations that align with increasingly abstract speech units.
\item Downsampling to lower resolutions within $\mrhubert$ does not improve downstream performance but improves computational efficiency.
\item Improved downstream performance of $\mrhubert$ over $\hubert$ is primarily due to the auxiliary loss located earlier in the network.
\end{itemize}

\begin{figure}
  \centering
  \includegraphics[scale=0.33]{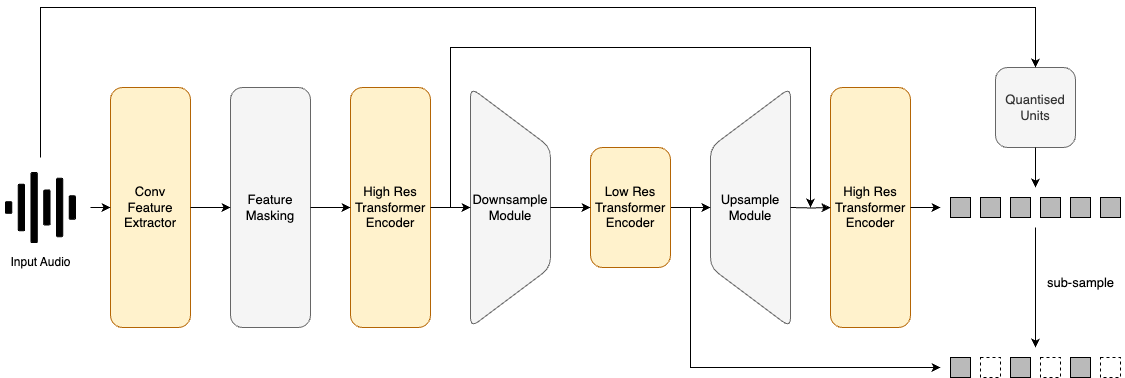}
  \caption{$\mrhubert$ framework which incorporates masked unit prediction at multiple resolutions.}
  \label{fig:MR-HuBERT}
\end{figure}

\section{Multi-Resolution HuBERT}
Hidden-Unit BERT ($\hubert$) \cite{hsu2021hubert} has established itself as the leading architecture for audio SSL models. For this reason, we focus here on Multi-Resolution HuBERT ($\mrhubert$) \cite{shi2024multiresolution, shi2023explorationhubert}, a model that aims to improve $\hubert$ through a multi-resolution architecture by introducing:
\begin{itemize}[noitemsep]
    \item \textbf{Downsample and upsample modules} between encoder blocks to process features at different resolutions (skip connections are applied to link encoders of the same resolution);\footnote{We note a potential error in the official implementation of $\mrhubert$ (see Appendix \ref{app:change-residual}).}
    \item \textbf{An auxiliary loss at low-resolutions}, applied at the end of each decoder and computed through a projection layer. Targets are formed by sub-sampling the base target stream.
\end{itemize}

We illustrate $\mrhubert$'s architecture in Fig. \ref{fig:MR-HuBERT}. In this paper, we analyse the layer-wise acoustic and linguistic information content of a series of ablations of the \texttt{MR-HuBERT-base} model\footnote{$\mrhubert$ models are downloaded from \href{https://github.com/facebookresearch/fairseq/blob/main/examples/mr\_hubert}{the Fairseq $\mrhubert$ page}.},
listed in Table \ref{tab:ablations} , and \texttt{HuBERT-base}\footnote{$\hubert$ models are downloaded from \href{https://github.com/facebookresearch/fairseq/blob/main/examples/hubert}{the Fairseq $\hubert$ page}.}. To do so, we employ methods used in previous representation analysis studies \cite{pasad2021wav2vec2, pasad2023models, pasad2024words}, such as Canonical Correlation Analysis (CCA) \cite{hotelling1992relations}, Mutual Information (MI) \cite{voita2019mutualinformation}, and spoken Semantic Textual Similarity (STS) \cite{conneau-kiela-2018-senteval}\footnote{We use the official implementation, available on the \href{https://github.com/ankitapasad/layerwise-analysis/}{Layerwise Analysis repository}.}. We also run Speech processing Universal PERformance Benchmark (SUPERB) \cite{yang2021superb} downstream tasks and analyse learnt layer weightings. We provide further details on our methodology in Appendix \ref{app:analysis_methods}.

\begin{table}[ht]
    \centering
    \begin{tabular}{ccccc}
        \toprule
        \textbf{Model} & \textbf{Resolutions (ms)} & \textbf{Layers} & \textbf{Downsampling} & \textbf{Auxiliary loss} \\
        \midrule
        \texttt{HuBERT-base} & 20 & 12 & \xmark & \xmark \\
        \texttt{MR-HuBERT-base}\footnotemark & 20, 40 & 4, 4, 4   & \cmark & \cmark \\
        $\texttt{MR-HuBERT B2-a}$    & 20, 40, 80 &  3, 2, 2, 2, 3  & \cmark & \cmark \\
        $\texttt{MR-HuBERT B2-b}$    & 20, 40, 80 &  2, 2, 4, 2, 2  & \cmark & \cmark \\
        $\texttt{MR-HuBERT } \bfoura$   & 20, 40 & 4, 4, 4  & \cmark & \xmark \\
        $\texttt{MR-HuBERT } \bfivea$   & 20 & 4, 4, 4  & \xmark & \cmark \\
        \bottomrule
    \end{tabular}
    \vspace{0.1cm}
    \caption{$\hubert$ and $\mrhubert$ models used for analysis in this work. The total number of layers is the same for all models. In $\btwoa$ and $\btwob$ a third resolution is introduced. $\bfoura$ is only trained on a single loss. $\bfivea$ has a single resolution but retains an auxiliary loss.}
    \label{tab:ablations}
\end{table}

\footnotetext{\texttt{MR-HuBERT-base} refers to the mono-base model on the Fairseq $\mrhubert$ page.}

\section{Findings}
\begin{figure}
\centering
\begin{subfigure}[b]{.435\textwidth}
    \centering
    \includegraphics[width=1\linewidth]{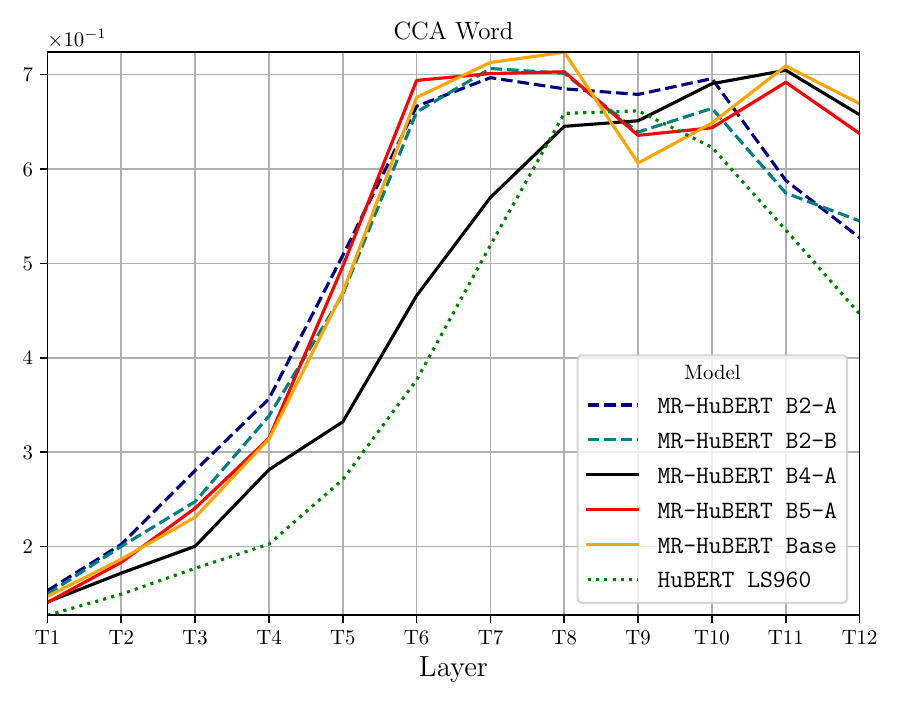}
    \caption{CCA word-level similarity.}
    \label{fig:cca_word}
\end{subfigure}
\hfill
\begin{subfigure}[b]{.535\textwidth}
    \centering
    \begin{subfigure}[t]{0.325\textwidth}
        \centering
        \includegraphics[width=\textwidth]{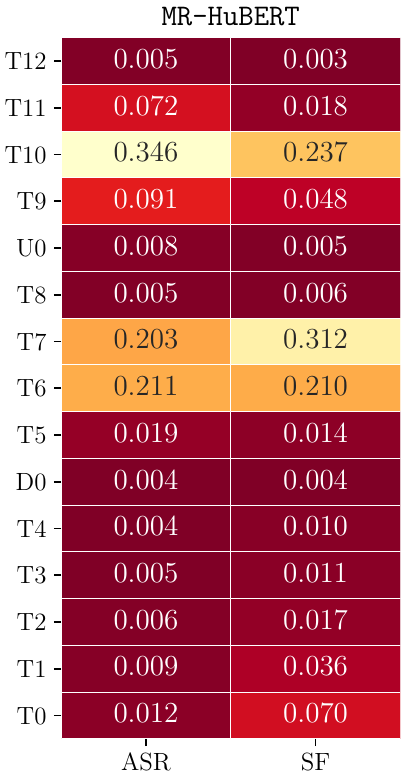}
    \end{subfigure}
    \hfill
    \begin{subfigure}[t]{0.325\textwidth}
        \centering
        \includegraphics[width=\textwidth]{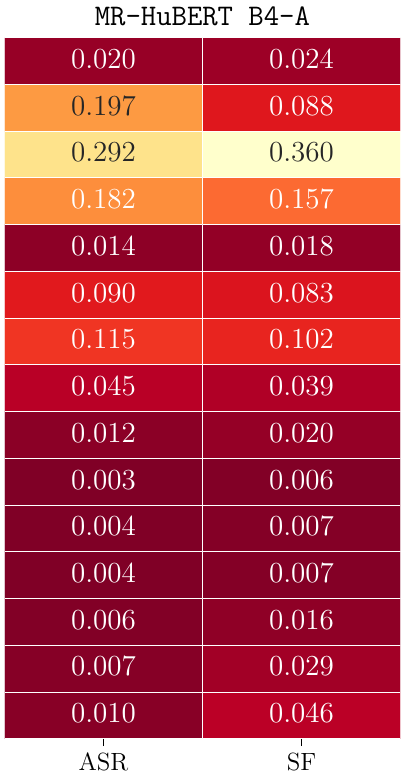}
    \end{subfigure}
    \hfill
    \begin{subfigure}[t]{0.325\textwidth}
        \centering
        \includegraphics[width=\textwidth]{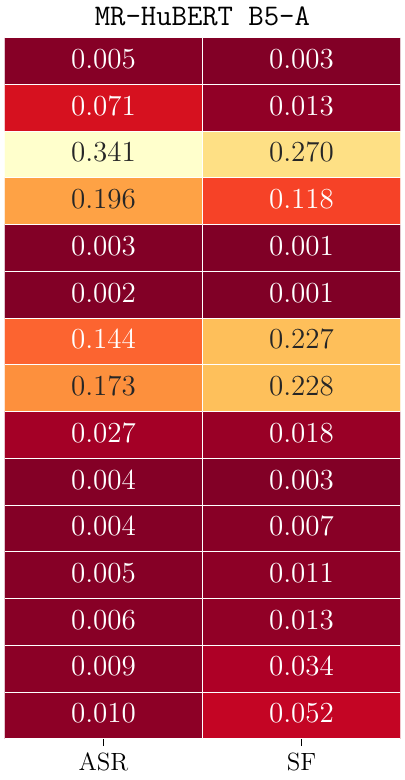}
    \end{subfigure}
    \caption{SUPERB layer importance-weightings.}
    \label{fig:superb_weightings}
\end{subfigure}
\caption{Impact of auxiliary loss, downsampling and added resolutions on information content and importance in downstream performance. 
Fig. \ref{fig:cca_word} shows CCA scores for $\hubert$ and multiple $\mrhubert$ variants. 
Comparing these models, we see that the auxiliary loss is the primary factor in increasing the word level information in earlier layers. 
Fig. \ref{fig:superb_weightings} shows SUPERB weights for the ASR and SF tasks, 
and again shows that the auxiliary loss is responsible for middle layers being useful for downstream tasks. \protect\footnotemark{} }
\label{fig:main_fig}
\end{figure}

\footnotetext{To explain the difference in number of layers between Figs. \ref{fig:cca_word} and \ref{fig:superb_weightings}: as discussed in appendix D.4 of \cite{shi2024multiresolution}, $\mrhubert$ encompasses transformer layers as well as outputs of the sampling modules, so a two-resolution $\mrhubert$ adds two layers, denoted by D0 and U0 in Fig. \ref{fig:superb_weightings}. Additionally, Fig. \ref{fig:cca_word} does not include the layer before the first transformer layer, denoted by T0.}

\subsection{Lower-resolution layers fail to capture abstract speech units}

In Fig. \ref{fig:cca_word}, we show the layerwise word-level CCA values of $\hubert$ and $\mrhubert$ models. We observe most $\mrhubert$ models feature two peaks: one near the end of the network (a feature of $\hubert$ models generally), and another near the middle (a feature of other SSL models \cite{pasad2023models}). Notably, downsampling alone does not change this pattern. We see no difference in word-level CCA between $\texttt{MR-HuBERT-base}$ (two-resolutions, downsampling) and $\bfivea$ (single resolution, no downsampling) nor do we see differences between $\mrhubert$ and three-resolution ablations (\texttt{B2-a} and \texttt{B2-b}). 

This pattern is consistent across other word-level measures (see Fig. \ref{fig:metrics_base} and \ref{fig:metrics_large} for further plots on other metrics and model sizes) as well as different speech units. We see no increase in learned word (Fig. \ref{fig:main_fig}),  phone or semantic (Fig. \ref{fig:metrics_base}) information in $\mrhubert$ when downsampling to various degrees ($\texttt{MR-HuBERT-base}$, \btwoa , \btwob, \texttt{B4-a}) compared to not downsampling at all ($\bfivea$). 

These results suggest that downsampling at these rates does not affect the information content of the representations learned. Most importantly, downsampling does not align with more abstract speech units. Whilst we find that middle layers of $\mrhubert$ are more heavily associated with more abstract information such as words, this appears to be independent of downsampling ($\bfivea$) and unaffected by the resolution at which these layers operate, see e.g. Fig. \ref{fig:superb_weightings}.

\subsection{Down-sampling is only helpful from an efficiency perspective}

Not only does downsampling in $\mrhubert$ not enhance the information content of representations, it also does not improve downstream performance. Table \ref{tab:superb_results} shows that performance in downstream tasks is hardly affected when downsampling is removed ($\bfivea$). This suggests downsampling is not responsible for improvements seen in $\mrhubert$ \cite{shi2024multiresolution}. Nevertheless, it is important to note that downsampling is still useful for improving model inference speed and training time. The current downsampling methods, however, appear too limited in scope to effectively capture broader linguistic units which naturally vary across time scales up to 50 times larger \cite{chen2023speechformer++}. This suggests more aggressive, context-aware downsampling techniques \cite{nawrot2022hierarchicaltransformersefficientlanguage} could better capture higher-level speech information, leading to both improvements in downstream performance as well as further efficiency gains.

\subsection{The auxiliary loss improves downstream performance}

In contrast, the removal of the auxiliary loss impacts our analysis significantly. We see worse performance of the $\bfoura$ model compared to $\texttt{MR-HuBERT-base}$ and $\bfivea$ on ASR tasks in Table \ref{tab:superb_results}. We also see clear differences in the content of the model representations in Fig. \ref{fig:main_fig} which could explain the observed difference in performance. In Fig. \ref{fig:cca_word}, we find the additional early peak typical of $\mrhubert$ entirely disappears when the auxiliary loss is removed ($\bfoura$; see also Fig \ref{fig:metrics_base}). Moreover, middle layers of the network are more useful for phonetic-based tasks, such as ASR and Slot Filling (SF), when the auxiliary loss ($\bfivea$) is present, independent of downsampling as shown in Fig. \ref{fig:superb_weightings}. We also find that $\bfoura$ results are closer to those of $\hubert$ than any other $\mrhubert$ model. This is the case for both ASR performance as shown in Table \ref{tab:superb_results}) as well as CCA scores as shown in \ref{fig:cca_word}\footnote{Remaining differences between $\bfoura$ and $\hubert$ may be due to $\mrhubert$'s extra training iteration \cite{shi2024multiresolution}. This may also explain why the peak scores for most CCA metrics are higher for $\mrhubert$ than $\hubert$.}. 

These results strongly suggest that the auxiliary loss is the key driver of downstream performance improvements \cite{shi2024multiresolution}. By encouraging the model to learn more diverse and relevant features at earlier layers, the auxiliary loss enhances the model’s ability to capture crucial phonetic and linguistic information. Notably, this loss mirrors the approach used in Deeply Supervised Nets \cite{lee2015deeplysupervisednets}, where early losses are thought to improve gradient flow and feature robustness. It may also act as a regulariser \cite{szegedy2016CVPR}, helping the model learn more stable, generalised representations by adding constraints during training — a benefit that could be especially important in low-resource settings like LibriSpeech. 

\begin{table}
    \small
    \centering
    \begin{tabular}{ |c|c|c|c|c|c|c|}
        \hline
        Model & ASR  & SF  &SE  & IC  & KS  & SD \\
        &(WER $\downarrow$)&(F1/CER $\uparrow/\downarrow$)&(STOI \cite{taal2010stoi}/PESQ \cite{rix2001pesq} $\uparrow/\uparrow$)&(Acc $\uparrow$)&(Acc $\uparrow$)&(Acc $\uparrow$)\\
        \hline
        $\texttt{HuBERT-base}^{+}$ &6.34&\textbf{89/23}&0.93/2.55&98.4&96.5&N/A\\
        \texttt{MR-HuBERT-base}   & 5.85    & 89/24 & 0.94/2.53 & \textbf{98.6} & 95.7 & 94.8\\
        \texttt{MR-HuBERT B4-a}        & 6.35    & 89/24 & 0.94/2.53 & 98.1 & \textbf{96.7} & \textbf{95.1}\\
        \texttt{MR-HuBERT B5-a}       & \textbf{5.82}    & 88/26 & \textbf{0.94/2.55} & 98.3 & 96.3 & 94.9\\
        \hline
    \end{tabular}
    \vspace{2mm}
    \caption{Performance on SUPERB downstream tasks with various upstream models based on $\mrhubert$. The results for \texttt{HuBERT-base}$^{+}$ are taken from \cite{shi2024multiresolution}.}
    \label{tab:superb_results}
\end{table}

\section{Conclusion}
In this study, we find that the improved downstream performance of $\mrhubert$ is primarily due to the auxiliary loss function, rather than downsampling in the multi-resolution architecture. Empirically, the auxiliary loss promotes better learning in intermediate layers, leading to superior downstream task performance. While downsampling enhances computational efficiency, it does not improve linguistic representations or downstream performance. Additionally, we find no evidence that lower-resolution layers capture more abstract speech information, highlighting the need for more effective unsupervised learning. This paper highlights the importance of analysing representation quality to gain deeper insights into how well multi-scale architectures capture different abstractions of speech information. We leave the exploration of improved architectures based on this analysis to future work.

\clearpage
\section{Acknowledgements}
The authors would like to thank Will Williams, John Hughes, Akis Kefalas and Ana Olssen for their valuable help in providing feedback and guidance on earlier drafts of this paper.

\bibliography{references} 

\newpage
\appendix

\section{Analysis methods} \label{app:analysis_methods}
In this section, we discuss the various metrics used to assess the acoustic and linguistic information present in the representations of different layers of a self-supervised model.

\subsection{Canonical correlation analysis}

Following previous layer-wise comparative studies \cite{pasad2021wav2vec2, pasad2023models}, we employ Projection Weighted Canonical Correlation Analysis (PWCCA), referred to throughout this paper as simply CCA, in order to correlate the model's internal representations with phonetic and word information and investigate how this varies across layers in the model. The internal representation for each word/phone is calculated by averaging the model's representations across the time steps corresponding to the span of that word/phone in the input sequence. Averaging across the time dimension effectively condenses the sequence information into a single vector representation for each word/phone, facilitating a more straightforward comparison of model behaviour across different layers. This process is then repeated to compare these internal representations against a range of external representations, capturing different linguistic and phonetic characteristics. Specifically, we perform comparisons using the following sets of representations: CCA mel (representations based on MFCCs to capture phonetic features), CCA phone (one-hot encoded phoneme embeddings), CCA word (one-hot encoded word embeddings), CCA glove (GloVe word embeddings \cite{pennington2014glove} to capture semantic similarity), CCA agwe (acoustically grounded word embeddings \cite{settle2019agwe} reflecting spoken word characteristics). For each of these, we follow \cite{pasad2023models} by using 7000 samples of words/phones from Librispeech.

\subsection{Mutual information}

Mutual information (MI) measures the information one random variable contains about another random variable. High MI is equivalent to a large reduction in uncertainty of one random variable given knowledge of the other, which implies dependence \cite{thomas2006elements}.

\begin{equation}
    \text{I}(X;Y) = \sum_{x \in \mathcal{X}} \sum_{y \in \mathcal{Y}} p(x,y) \log \left( \frac{p(x,y)}{p(x)p(y)} \right)
\end{equation}

We assess the dependence of phone and word labels on hidden representations in $\mrhubert$ as described in \cite{pasad2021wav2vec2, voita2019mutualinformation}. We first obtain averaged model features (as described above for CCA) which are then clustered using K-means to obtain a discrete distribution for MI analysis. Similarly to \cite{pasad2023models}, we cluster phone representations with $k=500$ and word representations with $k=5000$ centres.

\subsection{Spoken sentence-level semantic textual similarity}
The spoken Semantic Textual Similarity (STS) allows us to examine the extent to which SSL representations capture utterance-level semantic content \cite{merkx21_interspeech}. Following \cite{pasad2024words} we calculate Spearman’s $\rho$ correlation between annotated human judgments and the predicted similarity scores of utterance pairs. Sentence-level similarity scores are extracted by taking the cosine similarity between the mean-pooled representation of each utterance in a pair \cite{pasad2024words}.

\subsection{SUPERB}
\label{app:superb_definition}

Speech processing Universal PERformance Benchmark (SUPERB) \cite{yang2021superb} is a set of benchmarking resources to evaluate the performance of a shared model across a variety of speech processing tasks. We report on the following SUPERB downstream tasks to give results across a broad spectrum of speech-related tasks: Automatic Speech Recognition (ASR), Slot Filling (SF), Speech Enhancement (SE), Intent Classification (IC), Keyword Spotting (KS) and Speaker Diarisation (SD).

We examine downstream performance as well as the learned weightings inside the downstream adaptor for each layer in the pre-trained model to gain insights into where the most useful information is located for specific downstream tasks. When training downstream models, we use the default hyperparameters, including the learning rate. SUPERB learns a weighted average of the representations from the different layers from the self-supervised upstream model. Following previous work \citep{chang2021exploration, chen2022unispeech, hung22_interspeech, chen22g_interspeech, mlsuperb, lin2023utility, shi2023explorationhubert, otake2023parameter, chen2022wavlm}, we use these learned weightings to determine if certain layers contain significant information important to a specific downstream task and if so, which layers those are. 

\section{Layer-wise analysis of single and multi-resolution models} \label{app:layerwise_analysis}
\begin{figure}
\centering
\begin{subfigure}[t]{.48\textwidth}
    \centering
    \parbox[t]{\linewidth}{
        \textbf{A} \\[0.5ex]
        \includegraphics[width=\linewidth]{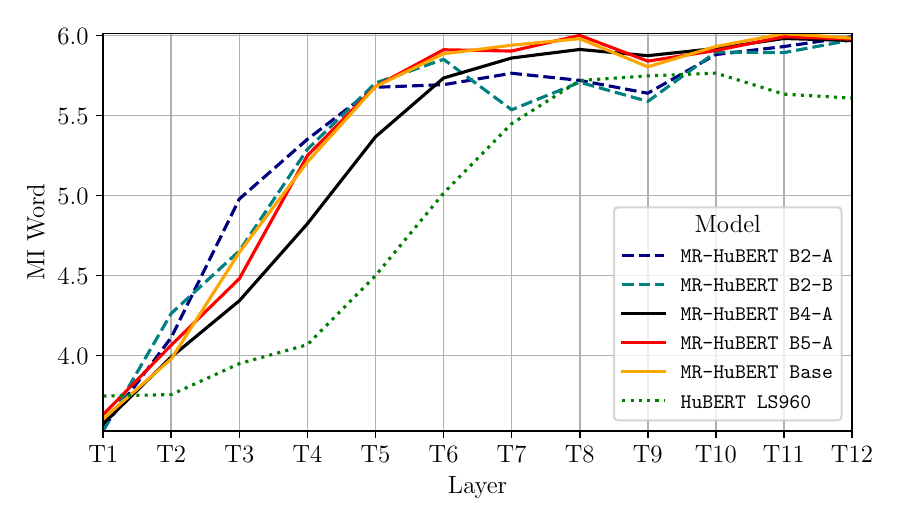}
    }
    \vspace{0ex}
    \label{fig:mi_word}
\end{subfigure}
\begin{subfigure}[t]{.48\textwidth}
    \centering
    \parbox[t]{\linewidth}{
        \textbf{B} \\[0.5ex]
        \includegraphics[width=\linewidth]{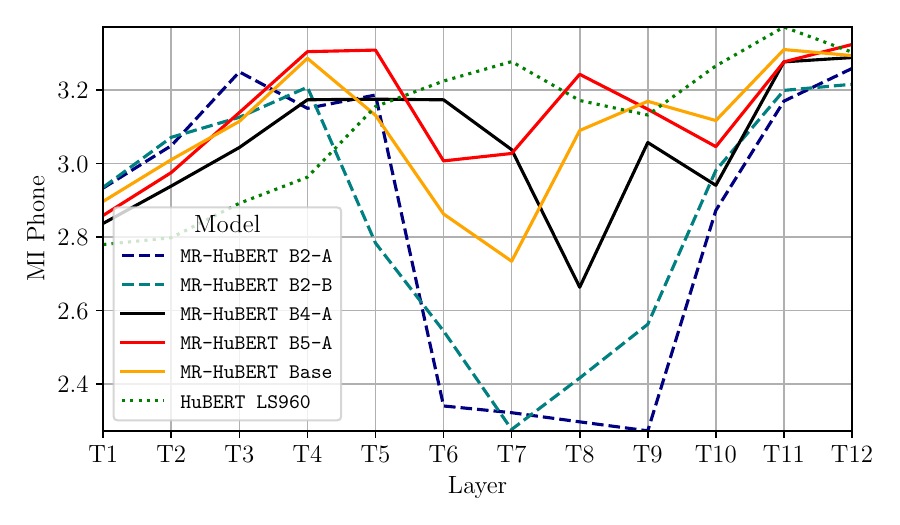}
    }
    \label{fig:mi_phone}
\end{subfigure}
\hfill
\begin{subfigure}[t]{.48\textwidth}
    \centering
    \parbox[t]{\linewidth}{
        \textbf{C} \\[0.5ex]
        \includegraphics[width=\linewidth]{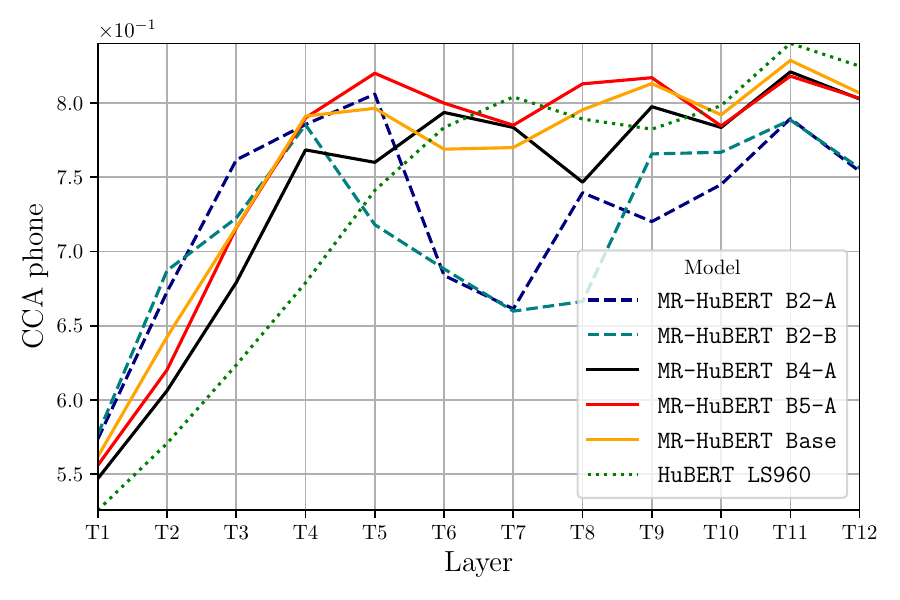}
    } 
    \label{fig:cca_phone}
\end{subfigure}%
\begin{subfigure}[t]{.48\textwidth}
    \centering
    \parbox[t]{\linewidth}{
        \textbf{D} \\[0.5ex]
        \includegraphics[width=\linewidth]{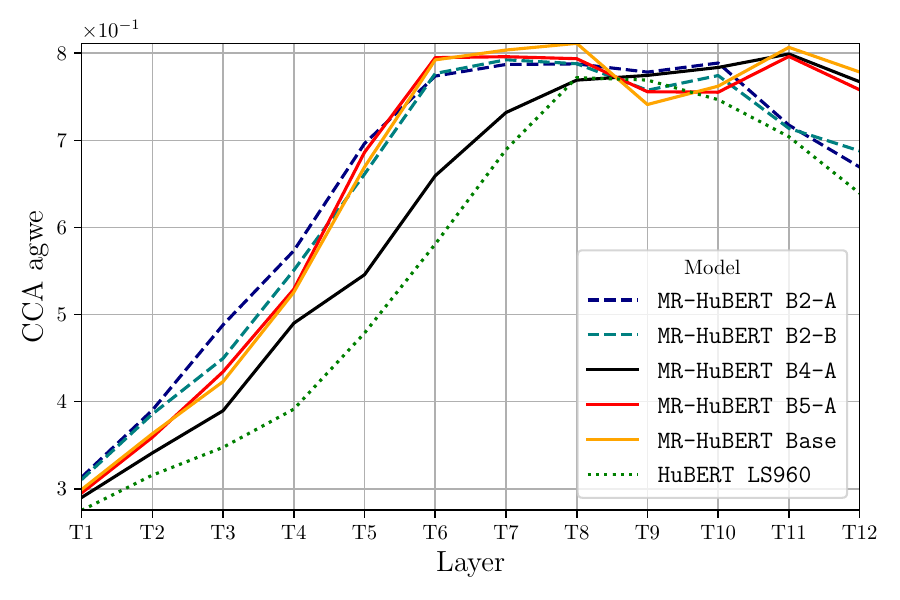}
    }
    \label{fig:cca_agwe}
\end{subfigure}
\hfill
\begin{subfigure}[t]{.48\textwidth}
    \centering
    \parbox[t]{\linewidth}{
        \textbf{E} \\[0.5ex]
        \includegraphics[width=\linewidth]{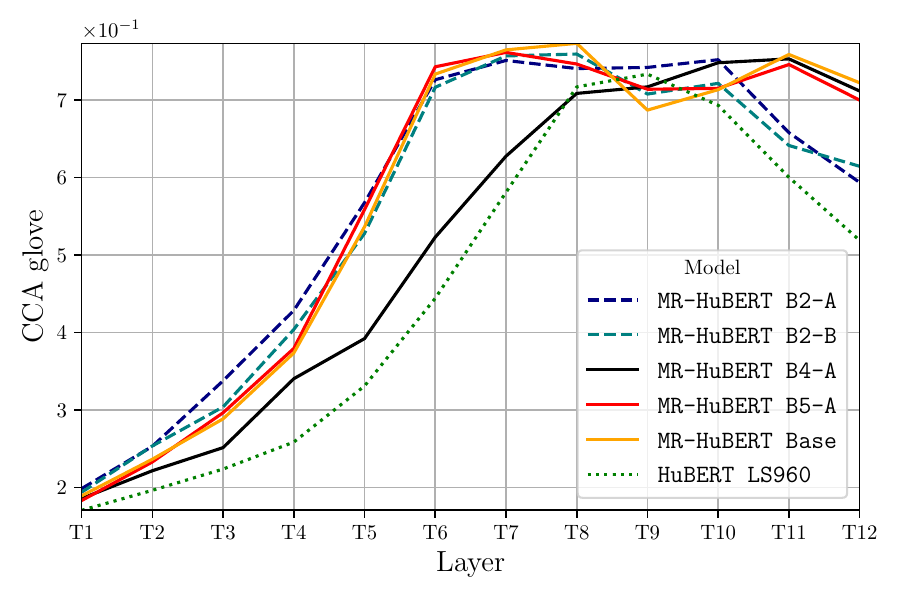}
    }
    \label{fig:cca_glove}
\end{subfigure}
\begin{subfigure}[t]{.48\textwidth}
    \centering
    \parbox[t]{\linewidth}{
        \textbf{F} \\[0.5ex]
        \includegraphics[width=\linewidth]{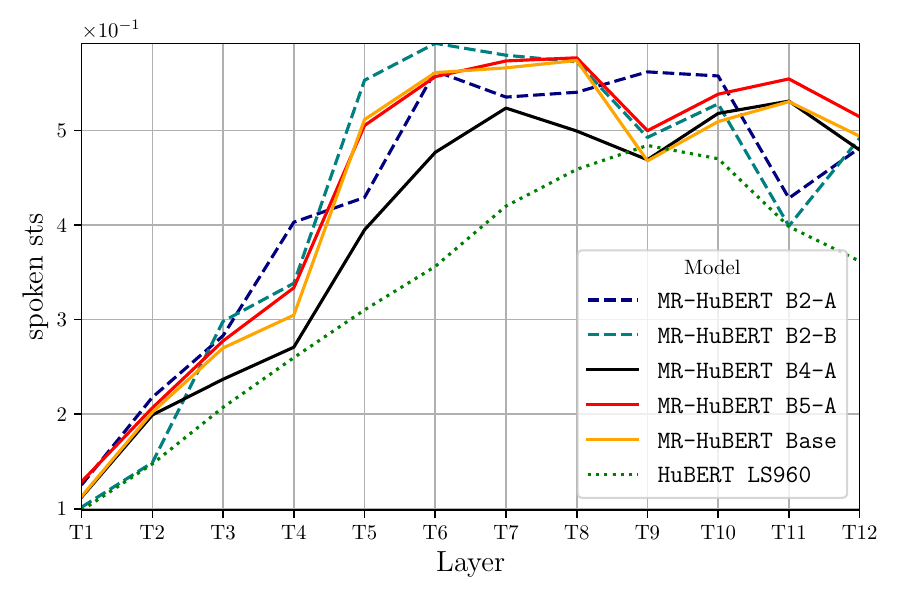}
    }
    \label{fig:spoken_sts}
\end{subfigure}
\caption{Layer-wise analyses of base models of $\mrhubert$ and $\hubert$  models. (A) MI scores between mean-pooled word-level representations and word identities. (B) MI scores between mean-pooled phone-level representations and phone identities. (C) CCA similarity between mean-pooled phone-level representations and phone identities (one-hot encoded). (D) CCA similarity between mean-pooled word-level representations and AGWE embeddings. (E) CCA similarity between mean-pooled word-level representations and GloVE embeddings. (F) Spearman’s $\rho$ correlation between annotated human judgments and cosine similarity of spoken utterance pairs.}
\label{fig:metrics_base}
\end{figure}

\begin{figure}[t]
\centering
\includegraphics[width=.9\textwidth]{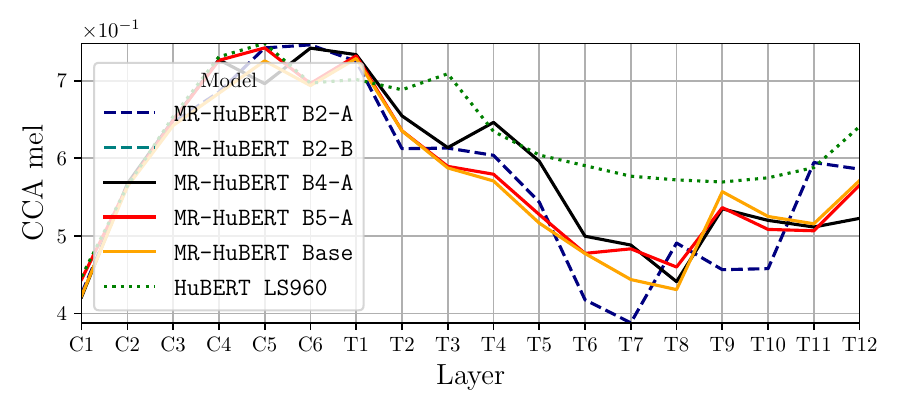} 
\caption{CCA similarity between frame-level representations and fbanks.}
\label{fig:cca_mel}
\end{figure}

In this section, we present the metric-specific results of the layer-wise analyses conducted on $\mrhubert$ ablations (see Table \ref{tab:ablations}) and $\hubert$ baselines. In addition to the findings reported in the paper, we observe in Fig. \ref{fig:cca_mel} that consistent with \cite{pasad2021wav2vec2}, the correlation between frame-level representations and fbanks increases with depth in the convolution layers of the feature extractor, but then decreases towards the mid-transformer layers for both $\hubert$ and $\mrhubert$ models. While there are no significant differences between two- and three-resolution models in frame, word, and sentence-level metrics, we do see a notable decrease in phone-level scores in the mid-layers of the three-resolution models (specifically \texttt{B2-a} and \texttt{B2-b} in panels B and C of Fig. \ref{fig:metrics_base}).

\section{Modifying the residual connection} \label{app:change-residual}
The diagram and equations from the $\mrhubert$ paper \cite{shi2024multiresolution} show that the residual for a given resolution is added before the decoder. However, the official implementation\footnote{https://github.com/facebookresearch/fairseq/blob/main/fairseq/models/multires\_hubert/multires\_hubert.py\#L783} contradicts this and adds the residual \textit{after} the decoder. We do not modify this in our experiments to retain consistency with the original results. However, we ran separate experiments which show an improvement in pre-train validation losses when the residual is added before the decoder (see Table \ref{tab:change-residual}).

\begin{table}
    \small
    \centering
    \begin{tabular}{ |c|c|c|}
        \hline
        Model & Final Train Loss  & Final Validation Loss\\
        \hline
        \texttt{post-residual (baseline)} &7.312&6.784\\
        \texttt{pre-residual}   &7.302&6.757\\
        \hline
    \end{tabular}
    \vspace{2mm}
    \caption{Effect on pre-train losses of altering the residual connection when added before the decoder.}
    \label{tab:change-residual}
\end{table}

These exploratory experiments used smaller $\mrhubert$ model sizes due to resource constraints. Models were trained for only 10\% of the usual 400k steps and the changes in architecture compared to $\mrhubert$-base are as follows: layers per encoder: 2, encoder embedding dim: 192, encoder feed-forward dim: 768.

\section{SUPERB layer weight analysis} \label{app:superb_layer_weights}
\begin{figure}
    \centering
    \begin{subfigure}[t]{.48\textwidth}
        \centering
        \includegraphics[width=\linewidth]{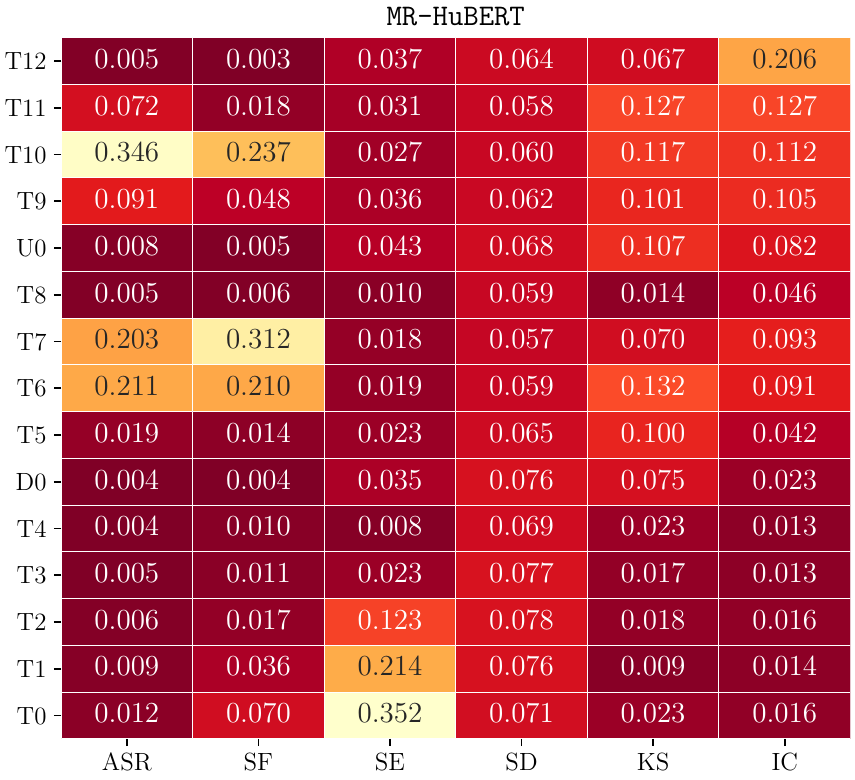}
        \caption{Layer weights for $\mrhubert$.}
        \label{fig:mono_base_weights}
    \end{subfigure}
\hfill
    \begin{subfigure}[t]{.48\textwidth}
        \centering
        \includegraphics[width=\linewidth]{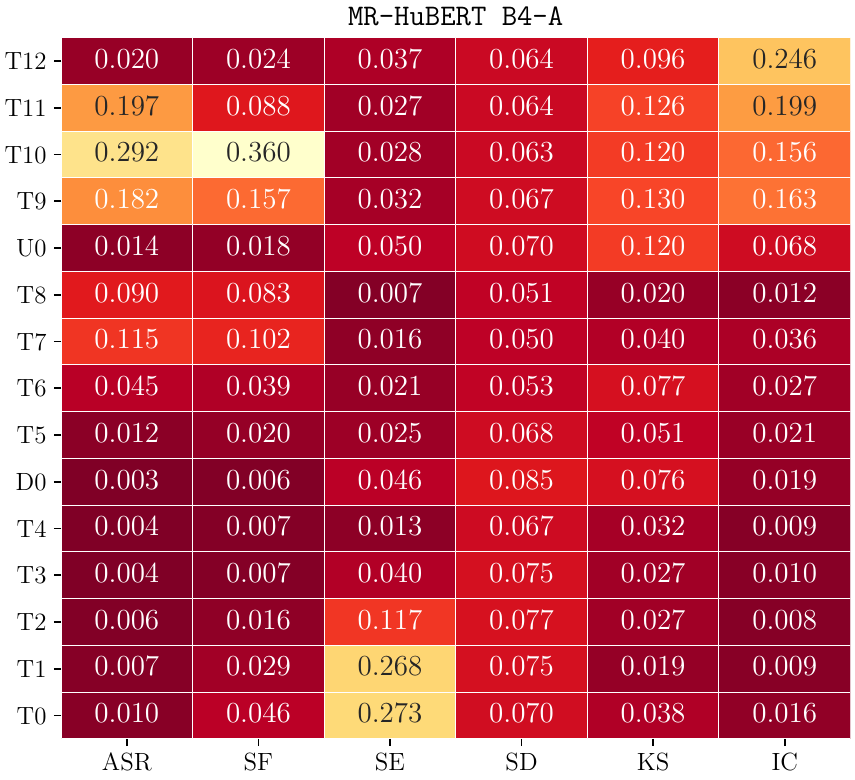}
        \caption{Layer weights for $\bfoura$.}
        \label{fig:b4a_weights}
    \end{subfigure}
    \begin{subfigure}[t]{.48\textwidth}
        \centering
        \includegraphics[width=\linewidth]{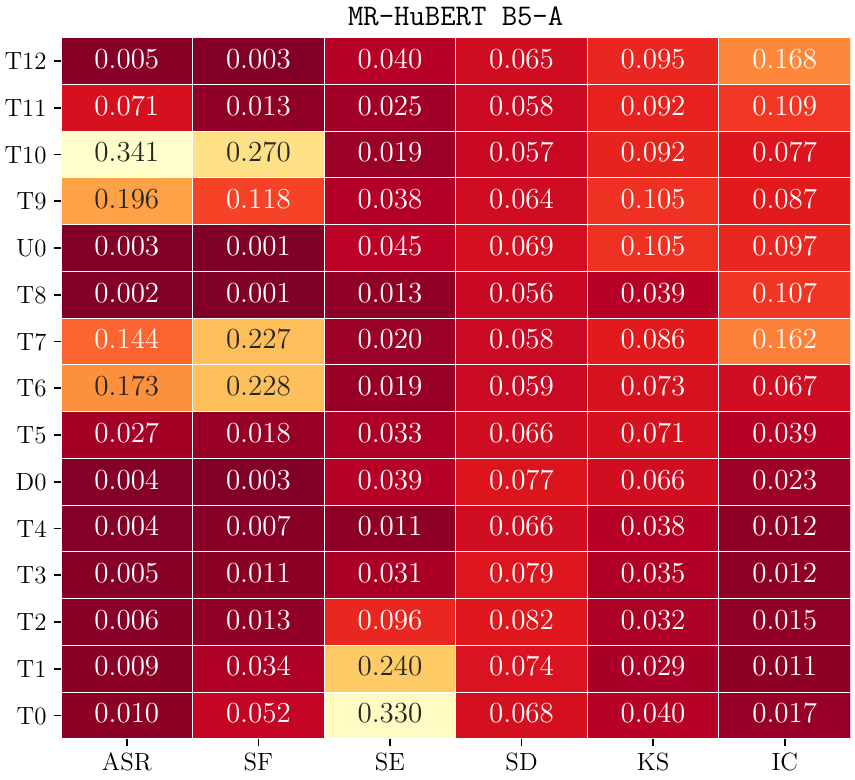}
        \caption{Layer weights for $\bfivea$.}
        \label{fig:b5a_weights}
    \end{subfigure}
    \caption{Layer importance-weightings for all SUPERB downstream tasks we study in this work, for $\mrhubert$ and two of its ablations.}
\end{figure}

Here, we provide further details on our layer weightings analysis of various SUPERB downstream tasks for a subset of the models listed in table \ref{tab:ablations}. 
We show a layer weight analysis for $\mrhubert$, $\bfoura$ and $\bfivea$ in Figs. \ref{fig:mono_base_weights}, \ref{fig:b4a_weights} and \ref{fig:b5a_weights} respectively to ablate the effects of auxiliary loss and the down- and upsampling modules further. The layer weightings in Fig. \ref{fig:mono_base_weights} support the same conclusions as in \cite{shi2024multiresolution}, e.g., $\mrhubert$ allocates over 40\% of its attention to low-resolution layers 8 and 9 for ASR. As discussed in the main text, this number decreases when downsampling is removed. We see a similar effect for the SF task, where focus is shifted away from the low-resolution encoder towards the second high-resolution encoder. The low-resolution $\mrhubert$ layers are associated to semantic context in the data and these results suggest that these semantics are pushed into the middle layers by training on the low-resolution loss and to a lesser extent by the downsampling. 

The SE task generally focuses on the early layers - at least 66\% of the weightings are assigned to the first three layers in all models. All the layers are used relatively evenly for SD and KS across all models. Weightings are slightly less concentrated towards the end of the network for $\bfivea$ compared to the other models.

\begin{figure}
\centering
\begin{subfigure}[t]{.48\textwidth}
    \centering
    \parbox[t]{\linewidth}{
        \textbf{A} \\[0.5ex]
        \includegraphics[width=\linewidth]{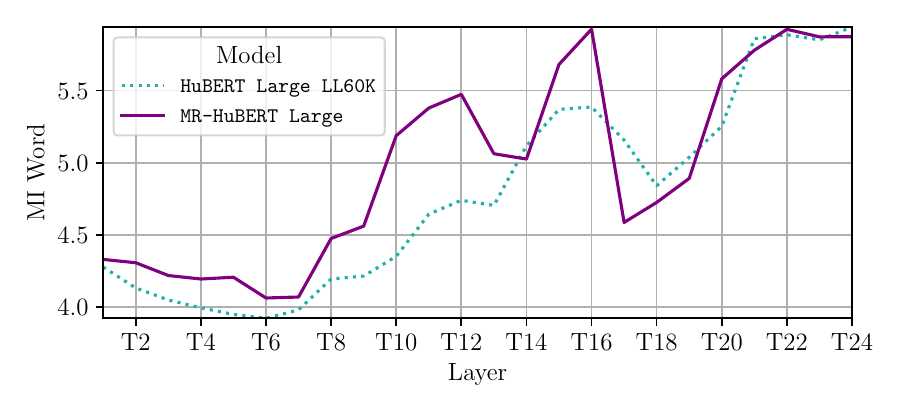}
    }
    \vspace{0ex}
    \label{fig:mi_large_word}
\end{subfigure}
\begin{subfigure}[t]{.48\textwidth}
    \centering
    \parbox[t]{\linewidth}{
        \textbf{B} \\[0.5ex]
        \includegraphics[width=\linewidth]{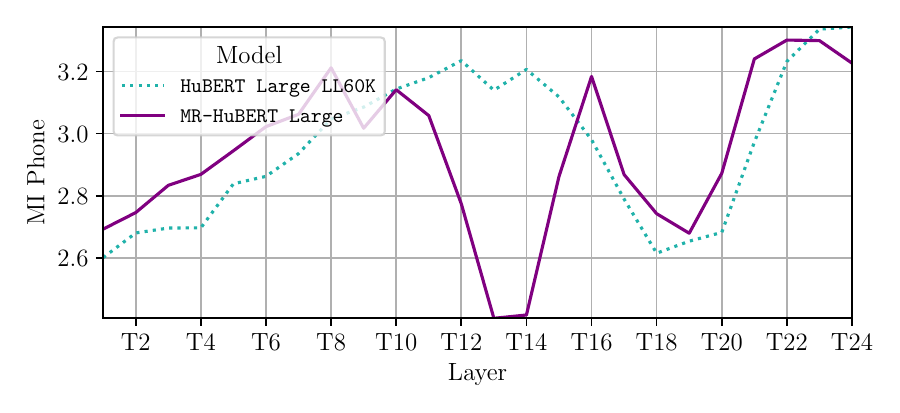}
    }
    \label{fig:mi_large_phone}
\end{subfigure}
\hfill
\begin{subfigure}[t]{.48\textwidth}
    \centering
    \parbox[t]{\linewidth}{
        \textbf{C} \\[0.5ex]
        \includegraphics[width=\linewidth]{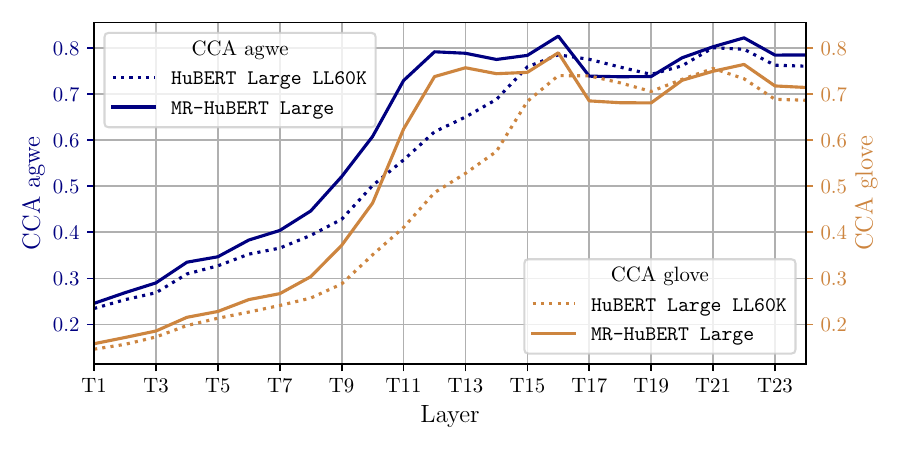}
    } 
    \label{fig:cca_large_agwq}
\end{subfigure}
\begin{subfigure}[t]{.48\textwidth}
    \centering
    \parbox[t]{\linewidth}{
        \textbf{D} \\[0.5ex]
        \includegraphics[width=\linewidth]{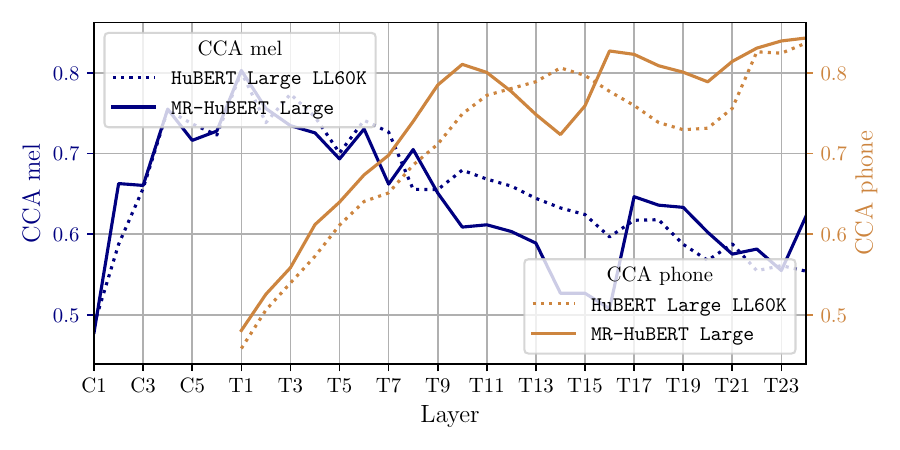}
    }
    \label{fig:cca_large_agwe}
\end{subfigure}
\hfill
\begin{subfigure}[t]{.48\textwidth}
    \centering
    \parbox[t]{\linewidth}{
        \textbf{E} \\[0.5ex]
        \includegraphics[width=\linewidth]{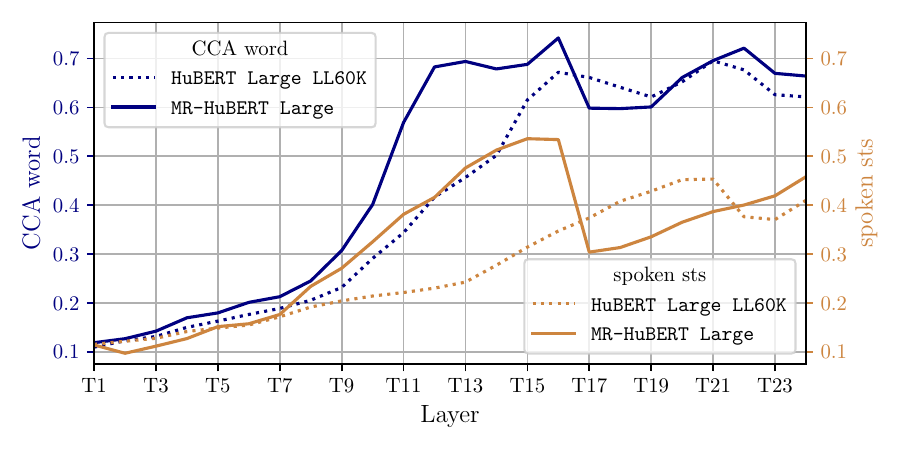}
    }
    \label{fig:cca_large_word}
\end{subfigure}
\caption{Layer-wise analyses of large models of $\mrhubert$ and $\hubert$  models. (A) MI scores between mean-pooled word-level representations and word identities (one-hot encoded). (B) MI scores between mean-pooled phone-level representations and phone identities (one-hot encoded). (C) CCA similarity between mean-pooled word-level representations and AGWE embeddings and GloVE embeddings. (D) CCA similarity between mean-pooled frame-level representations and fbanks as well as phone-level representations and phone identities (one-hot encoded). (E) CCA similarity between mean-pooled word-level representations and word identities (one-hot encoded) as well as Spearman’s $\rho$ correlation between annotated human judgments and cosine similarity of spoken utterance pairs.}
\label{fig:metrics_large}
\end{figure}

\end{document}